\documentclass{article}
\usepackage[utf8]{inputenc}
\usepackage[margin = 3cm]{geometry}
\usepackage{graphicx}
\usepackage{amsmath}
\usepackage{amssymb}
\usepackage{amsfonts}
\usepackage{tikz}
\usetikzlibrary{calc}
\usetikzlibrary{positioning}
\usetikzlibrary{arrows}

\usepackage[numbers]{natbib}

\usepackage{lmodern}

\usepackage{ifxetex,ifluatex}
\usepackage{fixltx2e} 
\ifnum 0\ifxetex 1\fi\ifluatex 1\fi=0 
  \usepackage[T1]{fontenc}
  \usepackage[utf8]{inputenc}
\else 
  \ifxetex
    \usepackage{mathspec}
  \else
    \usepackage{fontspec}
  \fi
  \defaultfontfeatures{Ligatures=TeX,Scale=MatchLowercase}
\fi
\IfFileExists{upquote.sty}{\usepackage{upquote}}{}
\IfFileExists{microtype.sty}{%
\usepackage{microtype}
\UseMicrotypeSet[protrusion]{basicmath} 
}{}

\usepackage{hyperref}
\hypersetup{unicode=true,
            pdftitle={Aim for clinical utility, not just predictive accuracy: Appendix},
            pdfauthor={Michael C Sachs and Arvid Sjölander and Erin E Gabriel},
            pdfborder={0 0 0},
            breaklinks=true}

\usepackage{color}
\usepackage{fancyvrb}

\DefineVerbatimEnvironment{Highlighting}{Verbatim}{commandchars=\\\{\}}
\usepackage{framed}
\definecolor{shadecolor}{RGB}{248,248,248}
\newenvironment{Shaded}{\begin{snugshade}}{\end{snugshade}}

\newcommand{\CommentTok}[1]{\textcolor[rgb]{0.56,0.35,0.01}{\textit{#1}}}

\newcommand{\ControlFlowTok}[1]{\textcolor[rgb]{0.13,0.29,0.53}{\textbf{#1}}}
\newcommand{\DataTypeTok}[1]{\textcolor[rgb]{0.13,0.29,0.53}{#1}}
\newcommand{\DecValTok}[1]{\textcolor[rgb]{0.00,0.00,0.81}{#1}}

\newcommand{\FloatTok}[1]{\textcolor[rgb]{0.00,0.00,0.81}{#1}}

\newcommand{\KeywordTok}[1]{\textcolor[rgb]{0.13,0.29,0.53}{\textbf{#1}}}
\newcommand{\NormalTok}[1]{#1}
\newcommand{\OperatorTok}[1]{\textcolor[rgb]{0.81,0.36,0.00}{\textbf{#1}}}
\newcommand{\OtherTok}[1]{\textcolor[rgb]{0.56,0.35,0.01}{#1}}

\newcommand{\StringTok}[1]{\textcolor[rgb]{0.31,0.60,0.02}{#1}}

\usepackage{graphicx,grffile}
\makeatletter
\def\maxwidth{\ifdim\Gin@nat@width>\linewidth\linewidth\else\Gin@nat@width\fi}
\def\maxheight{\ifdim\Gin@nat@height>\textheight\textheight\else\Gin@nat@height\fi}
\makeatother
\setkeys{Gin}{width=\maxwidth,height=\maxheight,keepaspectratio}
\IfFileExists{parskip.sty}{%
\usepackage{parskip}
}{
\setlength{\parindent}{0pt}
\setlength{\parskip}{6pt plus 2pt minus 1pt}
}
\ifx\paragraph\undefined\else
\let\oldparagraph\paragraph
\renewcommand{\paragraph}[1]{\oldparagraph{#1}\mbox{}}
\fi
\ifx\subparagraph\undefined\else
\let\oldsubparagraph\subparagraph
\renewcommand{\subparagraph}[1]{\oldsubparagraph{#1}\mbox{}}
\fi

\let\rmarkdownfootnote\footnote%
\def\footnote{\protect\rmarkdownfootnote}

\title{Aim for clinical utility, not just predictive accuracy}

\author{Michael C Sachs \and Arvid Sjölander \and Erin E Gabriel}

\begin{document}

\thispagestyle{empty}

\maketitle
\noindent
M.C. Sachs: Department of Medicine, Solna \\
              Eugeniahemmet, T2, Karolinska Universitetssjukhuset, \\
              171 76 Stockholm, Sweden \\
              Tel.: +46 08 517 761 42\\
              michael.sachs@ki.se \\
              \\
\noindent
           A. Sjölander and E.E. Gabriel:  \\
              Department of Medical Epidemiology and Biostatistics,\\
              Karolinska Institutet, \\
              171 65 Stockholm, Sweden

\noindent
\textbf{Running head} Aim for clinical utility, not just predictive accuracy \\
\textbf{Declaration} The authors declare that they have no conflict of interest. \\
\textbf{Word count} Abstract: 193; Main text: 3948 \\
\textbf{For consideration as} Original Research Article \\
\textbf{Sources of funding} The authors gratefully acknowledge financial support from the Swedish Research Council, Grant No. 2016-01267 and Starting Grant 2017-01898. \\
\textbf{Data and code} R Code used for simulated data and illustration is available as supplementary material included in the appendix.

\clearpage
\setcounter{page}{1}

\begin{center}
    \textbf{Aim for clinical utility, not just predictive accuracy}
\end{center}

\begin{abstract}
The predictions from an accurate prognostic model can be of great interest to patients and clinicians. When predictions are reported to individuals, they may decide to take action to improve their health or they may simply be comforted by the knowledge. However, if there is a clearly defined space of actions in the clinical context, a formal decision rule based on the prediction has the potential to have a much broader impact. Even if it is not the intended use of a developed prediction model, informal decision rules can often be found in practice. The use of a prediction-based decision rule should be formalized and compared to the standard of care in a randomized trial to assess its clinical utility, however, evidence is needed to motivate such a trial. We outline how observational data can be used to propose a decision rule based on a prognostic prediction model. We then propose a framework for emulating a prediction driven trial to evaluate the utility of a prediction-based decision rule in observational data. A split-sample structure can and should be used to develop the prognostic model, define the decision rule, and evaluate its clinical utility.

\end{abstract}

\section{Introduction}

Predictions of the prognosis of a patient given their current health state are prolific in medicine, as are the studies and tools to obtain such predictions. Risk scores, risk rankings, clinical outcome predictions, or \emph{prognostic prediction}, regardless of the name, are all attempting to predict with high accuracy the future outcome of a patient given current information. Despite the ubiquity of such predictions, the vast majority of the predictions and tools go unused in clinical practice \citep{collins2014external}. The lack of use may be due to the poor predictive accuracy of the predictions, but it may also be due to the lack of focus on the clinical context and the lack of actionable information provided by risk scores alone. Physicians want tools to help make treatment decisions, and to achieve this aim, one must clearly specify the clinical context and the intended use of a prognostic model at the outset of the study. If the intended use of a prognostic model is to drive treatment decisions, then it could be considered software as a medical device, and governing bodies in the US and Europe have made it clear that evidence of positive impact on patients, i.e., \emph{clinical validity} is required \citep{healthSoftwareMedicalDevice2019}. According to the FDA guidelines, a high degree of predictive accuracy is not necessary, as long as the product is analytically and clinically valid.

Prognostic information is inherently useful, allowing patients to cope with diagnosis and gain perspective. Knowledge of the individual predictions may lead to an action that improves the clinical outcome or quality of life of the patient. Taking actions in response to knowledge of a predicted future event is a prediction-based decision. A \emph{prediction-based decision rule} is a rule for taking an action in response to a prediction and it may be informal, unspecified, and varying across individuals, or it may be formalized as clinical guidelines for a population. These decisions may be driven by necessity, like hospital triage, or by the desire to lessen unpleasant side effects, like only treating aggressive cancers with toxic treatments. Regardless of the reasoning, in order for a prediction-based decision rule to be clinically useful, it is not sufficient that the prediction model be accurate. Instead, the use of the prediction-based decision rule should lead to improved clinical outcomes.

In settings where the set of all reasonable and possible actions in response to a prediction, the \emph{action space}, is well-specified, the development of a formal prediction-based decision rule can have a broad public health impact. As with all medical interventions, a prediction-based decision rule should be evaluated for \emph{clinical utility}, i.e., that using it for the intended purpose has a positive impact on patient clinical outcomes in comparison to the standard of care. The gold standard for generating evidence of clinical utility is a randomized controlled trial. While commonly understood for the evaluation of treatments, direct randomized comparisons of prediction-based decision rules to standard of care have received less attention \citep{steyerbergPrognosisResearchStrategy2013}.

Even when randomized trials evaluate prediction-based decision rules, they rarely directly assess the clinical utility of using the prediction-based decision rule in comparison to standard of care, instead simply looking for an interaction between treatment effect and prognosis. For example, the Oncotype DX risk score is a 21-gene based risk prediction model for distant recurrence of breast cancer that is currently being used in practice to determine whether chemotherapy should be used in addition to hormonal therapy \citep{kwa2017clinical}. After the risk score was developed for prognosis, the initial study motivating the use of the prediction to guide treatment was done retrospectively in a clinical trial that randomized subjects to compare hormonal therapy to hormonal therapy plus chemotherapy \citep{paik2006gene}. In that study, it was shown that there was a significant interaction between the risk score and treatment arm. This merely suggests that treatment effects differ by predicted prognosis, not that using the prediction-based decision rule is superior to the standard of care, which was to treat everyone with hormonal plus chemotherapy. The recent TailorX trial of Oncotype DX, used a stratified design in which low risk patients were given hormonal therapy alone, high risk patients were given hormonal plus chemotherapy, and medium risk patients were randomized to the two possible treatments \citep{sparano2018adjuvant}. The results showed that the hormonal therapy alone was not inferior to combination therapy in the medium group. Thus, even one of the most well known and used prediction-based decision rules, high Oncotype DX risk score for chemotherapy, which has had multiple randomized clinical trial comparisons, has not been directly compared to the standard of care for clinical superiority.

In the context of comparing the effectiveness of treatments, \citep{hernan2016using} has proposed and promoted the concept of emulating a target trial with observational data, and this concept has also been applied to assess the utility of screening \citep{garcia2017value}, and in other clinical settings \citep{caniglia2019emulating}. In this paper, we similarly propose to use observational data to emulate a target trial to assess the clinical utility of a prediction-based decision rule. Our proposal shares some key features with the emulated treatment trial, such as the clear specification of the eligibility, so we will focus on the special considerations and the additional components that are required for the development and evaluation of prediction-based decision rules. Our proposal focuses directly on the benefit to patients by assessment of clinical utility of a prediction-based decision rule, rather than less relevant measures of predictive accuracy.


To make things concrete, we will consider the setting of major abdominal surgery in Crohn's Disease (CD). Major abdominal surgery due to CD is considered a serious adverse outcome, and is responsible for high health care costs and decrease in quality of life in people with CD. Identifying individuals at high risk for surgery may allow for targeted use of early therapeutic interventions to offset this natural course. Several researchers have developed risk prediction models for CD-related surgery and complications \citep{guizzetti2017development, sachs2019ensemble}. There have been randomized studies to compare the efficacy of early combination therapy to the standard of care \citep{khanna2015early}, but there has been no attempt, to our knowledge, to determine whether prediction-based decisions are beneficial. Throughout this paper, we will illustrate our concepts using this example. We provide a glossary of terms in the first section of the Appendix.

\section{Developing a proposal for a prediction-based decision rule} \label{opt}

To formalize the optimization of the use of a prognostic prediction in medical decision making, and arrive at a proposal for a prediction-based decision rule that can then be evaluated for clinical utility, the action space needs to be specified. The prediction-based decision rule then can be viewed as a mapping of the prognostic model result to the action space. The best decision rule, among those that can be evaluated given your data, is obtained by finding the mapping that optimizes the expected clinical utility. The ideal way to do this is to build a \emph{utility function}, i.e., a mapping from the decision rule to a weighted set of the possible outcomes, containing all the possible benefits, adverse events, and costs of treatment, and optimize this over a set of groupings based on the prediction and the possible treatments. One potential method for doing this is suggested in \citep{vickersDecisionCurveAnalysis2006}, but there are numerous ways to arrive at a `best' proposal for a prediction-based decision rule.

Suppose we want to use the prediction model for risk of surgery in CD described in \citep{sachs2019ensemble} to help determine which treatment patients should receive. As demonstrated in the paper, the predictive accuracy of the model is adequate for prognostic prediction alone, but as we will outline, this accuracy may be more or less important based on the how the decision rule is developed. To make the problem feasible, we might only consider different cutoffs of the prognostic prediction for major abdominal surgery in CD to define high and low risk, rather than the full (infinite) space of all possible groupings. Additionally, we will only consider two possible interventions (the action space), to treat with mono therapy (thiopurines alone) or to treat with combination therapy (thiopurines plus biologics). The optimal decision rule could then be any combination of these treatments and the cutoff between high and low risk based on the predictions. Our utility function is simply the proportion of patients who undergo surgery within 5 years, which we want to minimize.


To further reduce the optimization problem, we can separate these two pieces. For example, we could use a predictiveness curve \citep{pepe2007integrating}, which can be used to describe the distribution of the outcome conditional on the prediction, to select the `best' cutoff to define high and low risk. In our example, after the cutoff between high and low risk has been determined, there would only be two possible combinations of therapy and predicted risk, (1) low risk gets thiopurines alone and high risk gets thiopurines plus biologics, or (2) high risk gets thiopurines alone and low risk gets thiopurines plus biologics. One could estimate the utility of both of these decision rules in an observational data set, as we outline in detail in the Supplementary Materials. However, in practice in our setting, as is often the case, expert opinion dictates that thiopurines plus biologics should not be given to low risk patients. Although this is not directly optimizing the decision rule, expert opinion still allows us to develop a proposal for the prediction-based decision rule that would be useful in practice.

Regardless of how it is performed, this optimization is something that can and should be done in the observational setting prior to the running of the clinical trial. In a clinical trial, only a small set of decision rules can be considered, limiting the ability to optimize the decision. Instead, in the observational setting, with adequate data, all decision rules within the plausible space can be considered and proper utility/loss functions can be used to account for more than just the primary outcome. As noted above, the solution to the general problem of who to treat with what is a much larger problem, with a long literature motivated by subgroup identification in clinical trials. Even when only considering a currently available predictive algorithm and a small set of therapies, the possible approaches to picking the `best' prediction-based decision rule are numerous. The important feature of any useful method of optimization or selection of a proposed prediction-based decision rule is that it be data driven, and therefore it will need to be externally validated. For this we suggest that a randomly split validation set be used to run the emulated target study as outlined in Section \ref{trial}.

\section{Evaluating a proposed prediction-based decision rule in an emulated trial} \label{trial}

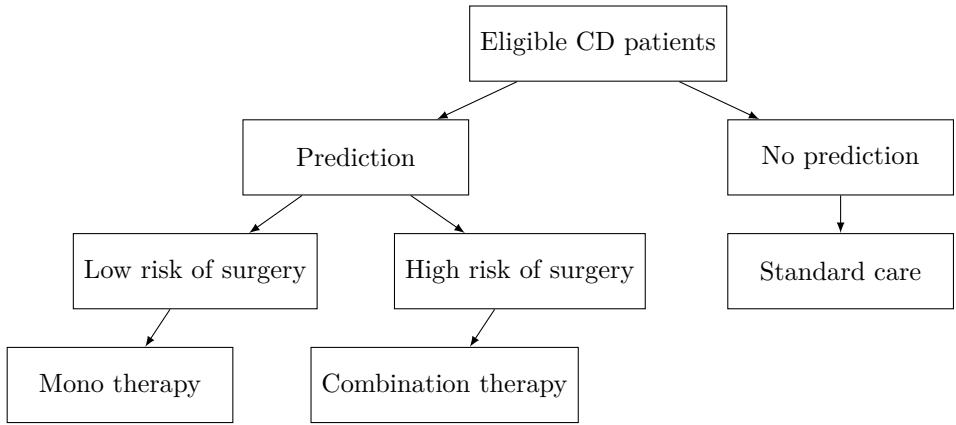
\begin{figure}[ht]
\begin{tikzpicture}[
    sharp corners=2pt,
    inner sep=4pt,
    node distance=.5cm and 0cm,
    >=latex]
\tikzstyle{my node}=[draw,minimum height=1cm,minimum width=3cm]
\node[my node] (elig){Eligible CD patients};
\node[my node,below left=of elig] (disease){Prediction};
\node[my node,below right=of elig] (nopred){No prediction};
\node[my node,below=of nopred] (nopredsoc){Standard care};
\node[my node,below left=of disease,xshift=1cm] (low){Low risk of surgery};
\node[my node,below right=of disease,xshift=-1cm] (high) {High risk of surgery};
\node[my node,below=of low,xshift=-1cm] (soc) {Mono therapy};
\node[my node,below=of high,xshift=-1cm] (esc) {Combination therapy};
\draw[->] (disease) -- (low);
\draw[->] (disease) -- (high);
\draw[->] (low) -- (soc);
\draw[->] (high) -- (esc);
\draw[->] (elig) -- (disease);
\draw[->] (elig) -- (nopred);
\draw[->] (nopred) -- (nopredsoc);
\end{tikzpicture}
\caption{Overview of the target prediction-model based trial in which patients are randomized to a prediction-based treatment decision rule versus standard of care. \label{trialdiag}}
\end{figure}

Suppose that we have a proposed decision rule for selecting between mono and combination therapy in patients recently diagnosed with Crohn’s disease, and before this rule is used in clinical practice, we wish to test it in some way for clinical utility. The highest quality evidence would be generated by a direct randomized comparison between the prediction-based decision rule and the standard treatment decision. Figure \ref{trialdiag} depicts this design for our running example in CD. Note that the only randomization is between using the prediction-based decision rule or the standard of care, which may or may not use other decision rules. Table \ref{tab:table1} summarizes the main components of a hypothetical target trial designed to compare the use of our prediction-based decision rule versus the standard of care.

In this study, we would obtain sufficient data on eligible study participants at the time of CD diagnosis to predict their risk of surgery within 5 years using the fixed prognostic model described above. Then, subjects would be randomized to the prediction-based decision arm, or the non-prediction-based decision arm (standard care). The decision process with the higher expected utility, in this case measured by the proportion undergoing surgery within 5 years, would be more beneficial to use in clinical practice. In the absence of this target randomized trial, we can use a large observational database to provide preliminary evidence or to motivate the funding of a randomized study by emulating this trial.

\begin{table}[h]
\caption{A summary of a target trial to evaluate the clinical utility of a proposed prediction-based decision rule in Crohn's disease.}
    \centering
    \begin{tabular}{p{4cm}|p{6.5cm}}
        Protocol Component & Description \\
        \hline
    Prediction model & Prediction for major abdominal surgery within 5 years based on clinical disease history, treatment history, and demographics. \\
    Prediction categories & Low risk and high risk of surgery. \\
    Action space & Mono versus combination therapy. \\
    Study design & Randomized comparison of prediction-based treatment decision to standard of care. High risk given combination therapy, low risk given monotherapy. \\
    Eligibility criteria & Adults with CD an no previous major abdominal surgery. \\
    Outcome & Major abdominal surgery within 5 years, yes or no. \\
    Clinical utility measure & Proportion undergoing surgery within 5 years. \\
    Analysis plan & Comparison of the clinical utility measure in the prediction-driven group to the non-prediction-driven group. \\
    \end{tabular}

    \label{tab:table1}
\end{table}

\citep{hernan2016using} proposes a framework for using modern methods of causal inference in observational data to emulate a clinical trial for comparing treatments. Just as \citep{hernan2016using} describes for the treatment comparison setting, we must define the eligibility criteria for our target trial and use similar criteria in our observational emulation. The criteria for our example are given in Table \ref{tab:table1}. The eligibility criteria define our time zero, as subjects would be enrolled as soon as they meet the criteria. Not all subjects will start therapy of any type precisely on the day when the eligibility criteria are met, and failing to account for this time window from eligibility to treatment can potentially impact the feasibility of conducting the study. In addition, to avoid any survivor bias that this grace period may cause, subjects having the event within the grace period should be included in both arms.

In our example, time zero would need to be shortly after CD diagnosis, when the initial treatment decision is needed. In our emulated study, we would set the grace period to two weeks, allowing subjects assigned to mono or combination therapy to be considered given those therapies if they were assigned them within two weeks of CD diagnosis. Subjects having surgery prior to two weeks post diagnosis with CD would need to be included in both arms regardless of treatment.

Using modern methods, and under certain assumptions, we can estimate the expected potential outcome for subjects using the prediction-based decision rule. However, unlike the trials outlined in \citep{hernan2016using}, the comparison is not between two treatments but between a prediction-based decision rule and the standard of care. The standard of care group will simply be the observed population, using the same eligibility criteria and the same time zero, but will include everyone. All subjects, regardless of what treatment they received, who meet the eligibility criteria will thus be included in the standard of care arm.

\subsection{Estimands and estimation}
For simplicity, let $Y_i=1$ indicate subject $i$ having surgery within 5 years of CD diagnosis, and 0 otherwise. Let the two treatment options in our example be labeled $A$ for mono therapy and $B$ for combination therapy. The comparison of interest in a randomized clinical trial run as in Figure \ref{trialdiag} is the proportion of subjects that underwent surgery within 5 years in the arm that used the prediction to determine therapy compared to the proportion of subjects that underwent surgery within five years that were randomized to standard of care.

The intention to treat estimand is \[E\{Y_i(\mbox{used prediction})\}-E\{Y_i(\mbox{standard of care})\},\] where $Y_i(\mbox{used prediction})$ is the potential outcome for subject $i$ under use of the prediction, the outcome under the use of prediction regardless of what subject $i$ was actually assigned to, and $Y_i(\mbox{standard of care})$ is their potential outcome under standard of care. Under standard assumptions in a randomized clinical trial, $E\{Y_i(\mbox{used prediction})\}-E\{Y_i(\mbox{standard of care})\}$ can be unbiasedly estimated by using the observed conditional outcomes $E\{Y_i|\mbox{used prediction}\}-E\{Y_i|\mbox{standard of care}\}$.

In our observational setting, however, we cannot equate $E\{Y_i(\mbox{used prediction})\}$ with $E\{Y_i|\mbox{used prediction}\}$ because individuals were not randomized to use the prediction-based decision rule. Instead, we can consider what makes up this expectation, given the deterministic treatment rule. The estimand $E\{Y_i(\mbox{used prediction})\}$ in the trial can be decomposed as
\begin{equation*}
Pr\{\mbox{low risk}\}*E\{Y_i(A)|\mbox{low risk}\} + Pr\{\mbox{high risk}\}*E\{Y_i(B)|\mbox{high risk}\}.
\end{equation*}
\sloppy We can easily estimate both $Pr\{\mbox{low risk}\}$ and $Pr\{\mbox{high risk}\}$ provided our observational sample is a simple random sample of the population of interest. The expectations $E\{Y_i(A)|\mbox{low risk}\}$ and $E\{Y_i(B)|\mbox{high risk}\}$ are conditional means of potential outcomes and thus require that we account for any confounders between therapy $A$ and the outcome within the low risk group, as well as any confounders between therapy $B$ and the outcome in the high risk group.

One way to estimate these potential quantities is through g-computation \citep{naimi2017introduction}. Considering $E\{Y_i(A)|\mbox{low risk}\}$ first, subset to those patients classified as low risk. Then specify and estimate a regression model, often referred to as the Q model, for the mean outcome conditional on the observed treatment $Z_i$ and observed covariates $C_i$ using this subgroup:
$$
E(Y_i | Z_i, C_i, \mbox{ low risk}) = g(\beta; Z_i, C_i).
$$
Then, predicted potential outcomes for each subject are obtained by setting $Z_i = A$ for each subject, combining with their observed covariates and the estimated regression coefficients, and obtaining a prediction $\hat{E}(Y_i | Z_i = A, C_i) = g(\hat{\beta}; A, C_i)$. Our estimated mean of the potential outcomes is the average of these predictions. This estimate relies on several assumptions:
\begin{enumerate}
    \item Positivity of treatment assignment: Within each subgroup defined by the covariates $C_i$, there must be a positive probability of receiving treatment $A$.
    \item No unmeasured confounding. All confounders of the effect of treatment on the outcome are measured.
    \item Correct specification of the Q model. The model above is correctly specified in terms of the treatment and covariates.
\end{enumerate}
Under these assumptions, our suggested estimated potential outcome means are consistent for the true potential outcome means. The argument applies equally to the estimation of $E\{Y_i(B)|\mbox{high risk}\}$ and other similar quantities.


The final piece of the desired estimand, $E\{Y(\mbox{standard of care})\}$, is easier to estimate and requires no additional assumptions. As all subjects in the observational data set that meet the eligibility criteria received the standard of care, this can simply be estimated by calculating the proportion of subjects that underwent surgery within five years of diagnosis, $E\{Y_i\}$.

We are glossing over some of the details regarding estimation in these cases, and thus we will not outline how inference should be undertaken, in general. However, it should be noted that although the subjects contributing to the estimation of $E\{Y(A)|\mbox{low risk}\}$ and $E\{Y(B)|\mbox{high risk}\}$ will be different, all subjects will contribute to the estimation of $E\{Y(\mbox{standard of care})\}$, and this must be accounted for in the variance estimate of the final comparison, for example, with the nonparametric bootstrap. In the Supplementary Materials we work through a step-by-step example of our proposed framework to demonstrate how it works.



\begin{figure}[ht]
\centering
\includegraphics[width=1.05\textwidth]{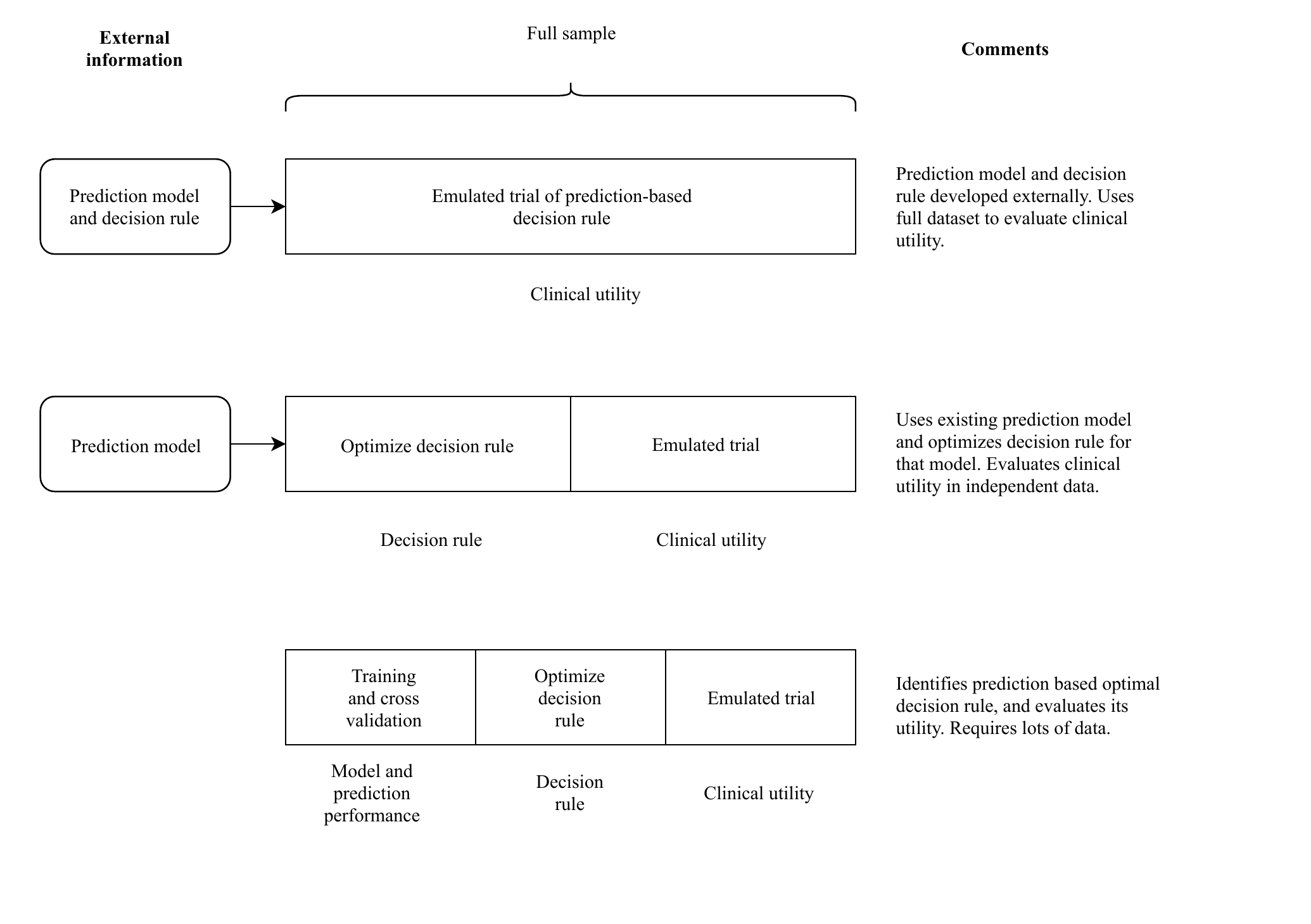}
\caption{Overview of strategies for split sample prediction model training and clinical utility assessment in observational settings. \label{splitdiag}}
\end{figure}

We have outlined our conceptual framework for developing a proposed prediction-based decision rule and evaluating its clinical utility in an emulated clinical trial. One can view this development and evaluation process of a prediction-based decision rule in three separate steps: first developing a prognostic model, then determining a proposed decision rule based on it, and finally evaluating the proposed decision rule for clinical utility. In a data-rich scenario, such as a large population based disease register, it may be possible to randomly split the dataset into a training sample, optimization sample, and clinical assessment validation sample. Figure \ref{splitdiag} illustrates the possible strategies for split sample training and optimization, and comments on their merits. If a good prognostic prediction model has already been developed, then it can be considered external information and only a two-way data split would be needed to do the second and third step. Likewise, if the prediction-based decision rule has already been proposed externally, all of the data can be used for the third step.

\section{Discussion}
If a prognostic prediction will be used to drive medical decisions, that prediction-based decision rule should be optimized and tested independently to determine whether it is worthwhile. To optimize and test a prediction-based decision rule, one must first define the space of possible actions that can be taken at the time or shortly after the prediction is made and propose a deterministic rule based on the predictions optimally. Then, evidence must be generated that the use of the decision rule based on the prediction model leads to improved clinical outcomes in comparison to the current standard of care. Often, the optimization in the development phase is overlooked, or is based on expert opinion without formal optimization, while the direct assessment of the utility of the prediction-based decision rule in comparison to the standard of care is even more rare.

We have outlined one approach to the development, optimization and evaluation of a prediction-based decision rule. This approach involves the optimization of a proposal for a prediction-based decision rule in observational data and the specification of the target prediction driven trial of interest, which can then be emulated in independent observational data, to formally evaluate the clinical utility of the optimized prediction-based decision rule.

The ideal procedure to develop an optimal decision rule based on baseline information may not involve prognostic prediction at all. Instead, one could directly optimize a decision rule for a utility function that quantifies all the benefits in terms of the relevant clinical outcomes, treatment decisions, as well as the costs. This decision rule would be a direct function of all available baseline information at the point a treatment decision was needed. There have been previous work on this in the context of identifying biomarkers for treatment selection in randomized trials \citep{dixon1991bayesian, morita2017bayesian}.

Using our proposed framework in observational data allows for optimization and assessment and therefore has the potential to improve current medical practice in a timely manner. Clearly, a randomized clinical trial evaluating a prediction-based decision rule would be ideal. We have discussed a simple two arm trial, but alternative randomized study designs may be useful in this context and their relative merits are summarized in \citep{micheel2012evolution}, Chapter 4. In lieu of this gold standard, an emulated trial should be used. As with all observational studies aiming to estimate causal contrasts, confounding is a major limitation. Prediction-based decision rules imply some special considerations regarding confounding. Prediction models are generally not designed with consideration of causal relationships, so colliders may be included in the models, see \citep{sjolander2009propensity}, and this has implications for the nature of the confounding that one needs to consider and control for. The precise conditions under which this is possible for different estimators needs to be explicitly investigated, and, like the assumption of no unmeasured confounding, it is likely that the assumptions that generate these conditions are untestable.

One practical limitation of our suggested approach is that the set of actions under consideration need to be observed in the sample to a sufficient degree to estimate the relevant conditional expectations (the positivity assumption). This may not be possible if the action space contains novel or unapproved treatments. For example, decision rules to treat with novel cancer drugs cannot be assessed unless those drugs are administered in practice. Other practical considerations, such as survivor and selection biases, which are inherent in retrospective studies, can be minimized by carefully planning studies with pre-specified protocols that mimic those used in randomized controlled trials \citep{hernan2016using, mcshane2013criteria}. This approach has been suggested and applied in comparative effectiveness research \citep{hernan2016specifying, garcia2017value, caniglia2019emulating} and we advocate for a similar approach. 

The optimization and utility assessment of a decision rule may depend on more than simply the clinical outcome. For instance, cost and quality of life considerations are important in many settings. Our proposed framework can apply such a utility function in both the optimization and evaluation phases, as long as the utility function is pre-specified and estimable using observed data. This is a concept that should be considered even in randomized clinical trials, but is likely easier to apply in register data, where many more measurements are available.

We have outlined the use of one target trial design for the evaluation of prediction-based decision rules, but as summarized in \citep{micheel2012evolution}, there are a variety of study designs that can be used to assess the value of predictions for guiding treatment decisions. Elucidation of this framework in alternative trial designs as well as the optimization step are of future research interest for the authors. Methods for identifying and optimizing biomarkers for treatment selection within clinical trials have a long history \citep{dixon1991bayesian}, and more recent methods incorporate modern machine learning approaches \citep{morita2017bayesian}. Applications of these methods in observations settings is possible in our proposed framework.


\bibliographystyle{aje}
\bibliography{references}

\clearpage

\section*{Appendix}

\section{Glossary of terminology}

\begin{itemize}
    \item[] \textbf{Prognostic prediction} A mathematical combination of multiple covariates/measurements that is used to calculate probabilities (or scores on other scales) of future events for individuals \citep{steyerbergPrognosisResearchStrategy2013}.
    \item[] \textbf{Action space} The set of possible decisions under consideration in a clinical context (e.g. to treat or not to treat) \citep{morita2017bayesian}.
    \item[] \textbf{Utility function} A mathematical quantification of the benefits, costs, and risks associated with decisions and future outcomes.
    \item[] \textbf{Prediction-based decision rule} A specified, logical and reproducible method for taking a prognostic prediction as input, and determining an element from the action space.
    \item[] \textbf{Clinical validity} The use of the model/decision rule achieves its intended purpose in the target population in the context of clinical care \citep{healthSoftwareMedicalDevice2019}.
    \item[] \textbf{Clinical utility} The use of the prediction-based decision rule in the target population and clinical context leads to improved outcomes for patients \emph{in comparison to the current standard of care.}
\end{itemize}

\section{Step-by-step Guide}

To  make  things  concrete,  we  will  consider  the  setting  of  major  abdominal  surgery  in Crohn’s Disease (CD). Major abdominal surgery due to CD is considered a serious adverse outcome, and is responsible for high health care costs and decrease in quality of life in people with CD. Identifying individuals at high risk for surgery may allow for targeted use of early therapeutic interventions to offset this natural course. We have simulated data to mimic a large national cohort of CD patients with clinical information, demographics, treatment history, and up to 10 years of follow up. Here we outline the steps that we will undertake to develop and evaluate a prediction based decision rule.

\subsection{Preparation and planning}
\begin{enumerate}
    \item Decide on eligibility criteria for the potential randomized clinical trial to assess a prediction based decision rule. This includes both patient characteristics and the time at which the decision will be made, e.g., within 2 weeks of diagnosis. Record this information in a draft protocol.
    \item Assemble a cohort of patients who meet the eligibility criteria. Define covariate data that is available before the time at which the decision is to be made, and then the clinical outcome of interest and the treatments that are observed at the time when the decision is to be made. Record this information in the draft protocol.
    \item Randomly allocate patients in the cohort into 3 groups: Model, decision rule, and clinical utility. If possible, keep the second group completely separate from your working environment until step 2.3 and the third group separate until step 2.4.
\end{enumerate}

\subsection{Development of prognostic model}
\begin{enumerate}
    \item In the first model subcohort, decide on the statistical techniques that will be used to develop prediction models. Record this information in a protocol.
    \item Apply the statistical techniques to develop a prediction model for the clinical outcome using the covariates that are available before the decision.
    \item In a cross-validation framework, assess the accuracy of the prediction model development process that was used in the previous step.
    \item Report the prediction model algorithm and its cross-validated performance. Test that the algorithm is reproducible and can be applied to new data in the same format.
\end{enumerate}

\subsection{Development of decision rule}
\begin{enumerate}
    \item In the decision rule subcohort, apply the prediction algorithm developed in the previous step. Examine the distribution of the predictions in this new cohort. Use predictiveness curves and clinical consultation to decide on cutoffs for high risk versus low risk. Record this decision making process and the results in a protocol.
    \item In consultation with an expert on the clinical context, and in view of the observed treatments in the sample, decide what the action space is. Record this in a draft protocol.
    \item Decide on the utility function that measures the utility of the treatment assignment. This is a function of the clinical outcome at a minimum, but may also incorporate quantifications of costs and risks associated with specific treatments. Record this in the protocol.
    \item Estimate the average utility for each combination of treatments in the action space and risk groups, accounting for confounders of the treatment outcome association using g-computation. The treatment with the highest average utility in each risk group defines the proposed prediction-based decision rule. Record this decision rule in the protocol.
\end{enumerate}

\subsection{Evaluation of the decision rule}
\begin{enumerate}
    \item In the clinical utility cohort, apply the prediction algorithm and cutoffs to obtain the classification into high risk and low risk.
    \item Specify the grace period as the time from eligibility to when a treatment was received (e.g., two weeks), and use this to define the observed treatment for each individual in the sample.
    \item In the subgroups of individuals classified as high and low risk separately, specify and fit a regression models for the outcome as a function of the observed treatment and observed confounders of the treatment-outcome association. Make sure to include relevant treatment-covariate interactions.
    \item Using the models estimated in step 3, obtain predictions for each subject by risk group using their observed covariates, and setting their treatment to the treatment prescribed by the proposed decision rule. These are the predicted potential outcomes.
    \item Term 1 of the estimand is the mean of the predicted potential outcomes times the proportion classified into each risk group and then added together:
    \begin{equation*}
    M_1 = Pr\{\mbox{low risk}\}*E\{\hat{Y_i}(A)|\mbox{low risk}\} + Pr\{\mbox{high risk}\}*E\{\hat{Y_i}(B)|\mbox{high risk}\}.
    \end{equation*}
    \item Term 2 of the estimand is the mean outcome in all subjects: $M_2 = E\{Y_i\}$.
    \item The estimated clinical utility is $M_1 - M_2$.
    \item To estimate the standard error, take a large number nonparametric bootstrap samples, and repeat steps 3-7 for each sample. The distribution of bootstrapped estimates can be used to calculate confidence intervals for the clinical utility.
\end{enumerate}

\section{R code with a simulated example}
\linespread{1}

\begin{Shaded}
\begin{Highlighting}[]
\KeywordTok{set.seed}\NormalTok{(}\DecValTok{20190826}\NormalTok{)}

\NormalTok{samp_data <-}\StringTok{ }\ControlFlowTok{function}\NormalTok{(}\DataTypeTok{n =} \DecValTok{2000}\NormalTok{) \{}

\NormalTok{    X <-}\StringTok{ }\KeywordTok{matrix}\NormalTok{(}\KeywordTok{rnorm}\NormalTok{(n }\OperatorTok{*}\StringTok{ }\DecValTok{3}\NormalTok{), }\DataTypeTok{ncol =} \DecValTok{3}\NormalTok{)}
\NormalTok{    trt.A <-}\StringTok{ }\KeywordTok{rbinom}\NormalTok{(n, }\DecValTok{1}\NormalTok{, }\DataTypeTok{p =} \KeywordTok{pnorm}\NormalTok{(X[, }\DecValTok{1}\NormalTok{] }\OperatorTok{*}\StringTok{ }\DecValTok{1} \OperatorTok{+}\StringTok{ }\NormalTok{X[, }\DecValTok{3}\NormalTok{] }\OperatorTok{*}\StringTok{ }\DecValTok{2}\NormalTok{))}

\NormalTok{    risk.p <-}\StringTok{ }\KeywordTok{pnorm}\NormalTok{(X[, }\DecValTok{1}\NormalTok{] }\OperatorTok{*}\StringTok{ }\FloatTok{.5} \OperatorTok{+}\StringTok{ }\NormalTok{X[, }\DecValTok{2}\NormalTok{] }\OperatorTok{*}\StringTok{ }\DecValTok{1} \OperatorTok{+}\StringTok{ }\NormalTok{X[, }\DecValTok{3}\NormalTok{] }\OperatorTok{*}\StringTok{ }\FloatTok{.75}\NormalTok{)}
\NormalTok{    highrisk <-}\StringTok{ }\NormalTok{risk.p }\OperatorTok{>}\StringTok{ }\FloatTok{.75}

\NormalTok{    true.Risk <-}\StringTok{ }\KeywordTok{rowSums}\NormalTok{(X)}
\NormalTok{    q3 <-}\StringTok{ }\KeywordTok{quantile}\NormalTok{(true.Risk, }\FloatTok{0.75}\NormalTok{)}
    \CommentTok{# potential outcomes: 1  = get trt.A if low risk, }
    \CommentTok{#                     0  = randomly assign trt.A according to X[, 1] and X[, 3]}
    \CommentTok{#                    in truth, trt.B is effective only at high risk}



\NormalTok{    p0 <-}\StringTok{ }\KeywordTok{pnorm}\NormalTok{(true.Risk }\OperatorTok{-}\StringTok{ }\KeywordTok{ifelse}\NormalTok{(true.Risk }\OperatorTok{>}\StringTok{ }\NormalTok{q3, }\DecValTok{2}\OperatorTok{*}\NormalTok{(}\DecValTok{1} \OperatorTok{-}\StringTok{ }\NormalTok{trt.A), }\DecValTok{0}\NormalTok{))}
\NormalTok{    Y0 <-}\StringTok{ }\KeywordTok{rbinom}\NormalTok{(n, }\DecValTok{1}\NormalTok{, }\DataTypeTok{p =}\NormalTok{ p0)}


\NormalTok{    trt.A.star <-}\StringTok{ }\KeywordTok{as.numeric}\NormalTok{(}\OperatorTok{!}\NormalTok{highrisk)}
\NormalTok{    p1 <-}\StringTok{ }\KeywordTok{pnorm}\NormalTok{(true.Risk }\OperatorTok{-}\StringTok{ }\KeywordTok{ifelse}\NormalTok{(true.Risk }\OperatorTok{>}\StringTok{ }\NormalTok{q3, }\DecValTok{2}\OperatorTok{*}\NormalTok{(}\DecValTok{1} \OperatorTok{-}\StringTok{ }\NormalTok{trt.A.star), }\DecValTok{0}\NormalTok{))}

\NormalTok{    Y1 <-}\StringTok{ }\KeywordTok{rbinom}\NormalTok{(n, }\DecValTok{1}\NormalTok{, }\DataTypeTok{p =}\NormalTok{ p1)}

    \KeywordTok{data.frame}\NormalTok{(X, trt.A, highrisk, }\DataTypeTok{Y =}\NormalTok{ Y0, }\DataTypeTok{Y1 =}\NormalTok{ Y1)}

\NormalTok{\}}

\NormalTok{estimate_utility <-}\StringTok{ }\ControlFlowTok{function}\NormalTok{(cucohort) \{}
    \CommentTok{## Step 3}

\NormalTok{    fit.high <-}\StringTok{ }\KeywordTok{glm}\NormalTok{(Y }\OperatorTok{~}\StringTok{ }\NormalTok{(X1 }\OperatorTok{+}\StringTok{ }\NormalTok{X2 }\OperatorTok{+}\StringTok{ }\NormalTok{X3) }\OperatorTok{*}\StringTok{ }\NormalTok{trt.A, }\DataTypeTok{family =} \StringTok{"binomial"}\NormalTok{,}
                    \DataTypeTok{data =} \KeywordTok{subset}\NormalTok{(cucohort, highrisk }\OperatorTok{==}\StringTok{ }\OtherTok{TRUE}\NormalTok{))}
\NormalTok{    fit.low <-}\StringTok{ }\KeywordTok{glm}\NormalTok{(Y }\OperatorTok{~}\StringTok{ }\NormalTok{(X1 }\OperatorTok{+}\StringTok{ }\NormalTok{X2 }\OperatorTok{+}\StringTok{ }\NormalTok{X3) }\OperatorTok{*}\StringTok{ }\NormalTok{trt.A, }\DataTypeTok{family =} \StringTok{"binomial"}\NormalTok{,}
                   \DataTypeTok{data =} \KeywordTok{subset}\NormalTok{(cucohort, highrisk }\OperatorTok{==}\StringTok{ }\OtherTok{FALSE}\NormalTok{))}

    \CommentTok{## Step 4}
    \CommentTok{## low risk gets treatment A}

\NormalTok{    newlow <-}\StringTok{ }\KeywordTok{subset}\NormalTok{(cucohort, highrisk }\OperatorTok{==}\StringTok{ }\OtherTok{FALSE}\NormalTok{)}
\NormalTok{    newlow}\OperatorTok{$}\NormalTok{trt.A <-}\StringTok{ }\DecValTok{1}
\NormalTok{    Yhat.A <-}\StringTok{ }\KeywordTok{predict}\NormalTok{(fit.low, }\DataTypeTok{newdata =}\NormalTok{ newlow, }\DataTypeTok{type =} \StringTok{"response"}\NormalTok{)}

\NormalTok{    newhigh <-}\StringTok{ }\KeywordTok{subset}\NormalTok{(cucohort, highrisk }\OperatorTok{==}\StringTok{ }\OtherTok{TRUE}\NormalTok{)}
\NormalTok{    newhigh}\OperatorTok{$}\NormalTok{trt.A <-}\StringTok{ }\DecValTok{0}
\NormalTok{    Yhat.B <-}\StringTok{ }\KeywordTok{predict}\NormalTok{(fit.high, }\DataTypeTok{newdata =}\NormalTok{ newhigh, }\DataTypeTok{type =} \StringTok{"response"}\NormalTok{)}

    \CommentTok{## Step 5}

\NormalTok{    M1 <-}\StringTok{ }\KeywordTok{mean}\NormalTok{(cucohort}\OperatorTok{$}\NormalTok{highrisk }\OperatorTok{==}\StringTok{ }\OtherTok{FALSE}\NormalTok{) }\OperatorTok{*}\StringTok{ }\KeywordTok{mean}\NormalTok{(Yhat.A) }\OperatorTok{+}\StringTok{ }
\StringTok{        }\KeywordTok{mean}\NormalTok{(cucohort}\OperatorTok{$}\NormalTok{highrisk }\OperatorTok{==}\StringTok{ }\OtherTok{TRUE}\NormalTok{) }\OperatorTok{*}\StringTok{ }\KeywordTok{mean}\NormalTok{(Yhat.B)}

    \CommentTok{## Step 6}
\NormalTok{    M2 <-}\StringTok{ }\KeywordTok{mean}\NormalTok{(cucohort}\OperatorTok{$}\NormalTok{Y)}

    \CommentTok{## Step 7}
\NormalTok{    M1 }\OperatorTok{-}\StringTok{ }\NormalTok{M2}
\NormalTok{\}}


\NormalTok{data <-}\StringTok{ }\KeywordTok{samp_data}\NormalTok{()}
\NormalTok{mainest <-}\StringTok{ }\KeywordTok{estimate_utility}\NormalTok{(data)}

\CommentTok{## Step 8}
\NormalTok{bootests <-}\StringTok{ }\KeywordTok{rep}\NormalTok{(}\OtherTok{NA}\NormalTok{, }\DecValTok{1000}\NormalTok{)}
\ControlFlowTok{for}\NormalTok{(i }\ControlFlowTok{in} \DecValTok{1}\OperatorTok{:}\KeywordTok{length}\NormalTok{(bootests))\{}
\NormalTok{    sampdex <-}\StringTok{ }\KeywordTok{sample}\NormalTok{(}\DecValTok{1}\OperatorTok{:}\KeywordTok{nrow}\NormalTok{(data), }\KeywordTok{nrow}\NormalTok{(data), }\DataTypeTok{replace =} \OtherTok{TRUE}\NormalTok{)}

\NormalTok{    bootests[i] <-}\StringTok{ }\KeywordTok{estimate_utility}\NormalTok{(data[sampdex, ])}
\NormalTok{\}}

\NormalTok{CI <-}\StringTok{ }\KeywordTok{quantile}\NormalTok{(bootests, }\KeywordTok{c}\NormalTok{(.}\DecValTok{025}\NormalTok{, }\FloatTok{0.975}\NormalTok{))}
\end{Highlighting}
\end{Shaded}

The estimated clinical utility of the decision rule is -0.11 95\% CI:
-0.17 to -0.05. This is interpreted as the estimated difference in the
proportion having the outcome comparing the prediction based decision
rule arm to the standard of care.

\end{document}